\documentclass[universe,article,accept,oneauthor,pdftex]{mdpi} 

\firstpage{1} 
\pubvolume{8}
\issuenum{2}
\articlenumber{69}
\pubyear{2022}
\copyrightyear{2022}
\externaleditor{{Academic} Editor: Panayiotis Stavrinos} 
\datereceived{16 December 2021} 
\dateaccepted{21 January 2022} 
\datepublished{23 January 2022} 
\hreflink{https://\linebreak doi.org/10.3390/universe8020069} % If needed use \linebreak
\usepackage{amssymb}
\Title{On the Conformal Frames in $f (R)$ Gravity}

\TitleCitation{On the Conformal Frames in $f (R)$ Gravity}

\Author{Yuri Shtanov $^{1, 2}$\orcidA{}}
\AuthorNames{Yuri Shtanov}

\AuthorCitation{Shtanov, Y.}
\address{%
	$^{1}$ \quad Bogolyubov Institute for Theoretical Physics,  Metrologichna St.\@ 14-b, 03143 Kiev, Ukraine; shtanov@bitp.kiev.ua\\  
	$^{2}$ \quad Department of Physics, Taras Shevchenko National University of Kiev, Volodymyrska St.\@ 60, \mbox{01033 Kiev, Ukraine}}%MDPI: Please add department.

\abstract{We discuss gravitational physics in the Jordan and Einstein frames of $f (R)$ gravity coupled to the Standard Model.  We elucidate the way in which the observed gravitational coupling arises in the Einstein frame for generic $f (R)$.  We point out that the effect of ``running units'' in the Einstein frame is related to the fact that the explicit and implicit quantum parameters of the Standard Model, such as the Higgs vacuum expectation value and the parameter $\Lambda_\text{QCD}$, are modified by the conformal transformation of the metric and matter fields and become scalaron-dependent. Considering the scalaron of $f (R)$ gravity describing dark matter, we show that the effect of running units  in this case is extremely weak, making two frames practically equivalent.}

\keyword{modified gravity}

\begin{document}
%%%%%%%%%%%%%%%%%%%%%%%%%%%%%%%%%%%%%%%%%%
%\setcounter{section}{-1} %% Remove this when starting to work on the template.

\section{Introduction}

Modification of the general relativity theory by considering the Lagrangian in the form of a nonlinear function $f (R)$ of the scalar curvature $R$ is, perhaps, the simplest one and has long been the subject of numerous studies and applications  (see~\cite{Sotiriou:2008rp, DeFelice:2010aj, Nojiri:2017ncd} for reviews). Compared to the general relativity theory, such $f (R)$ gravity contains one extra degree of freedom, which can be used for modelling a wide variety of phenomena, from the inflationary regime in the early universe~\cite{Starobinsky:1980te} to dark matter at later epochs~\cite{Nojiri:2008nt, Cembranos:2008gj, Corda:2011aa, Katsuragawa:2016yir, Yadav:2018llv, Parbin:2020bpp, Shtanov:2021uif}. 

The extra degree of freedom is most conveniently identified in the so-called Einstein frame of fields, where it becomes a separate scalar field, called the scalaron, while the remaining gravitational degrees of freedom are described by the general-relativistic action.  The existence of different conformal frames has long ago raised the issue of their physical equivalence~\cite{Dicke:1961gz}.  After some debates, this question, in principle, appears to have been resolved (see~\cite{Faraoni:2006fx}). The conformal frames are physically equivalent and describe the same observable phenomena if one carefully takes into account conformal transformation of all masses. Thus, in the Einstein frame, one then deals with so-called ``running units'', with all physical masses becoming scalaron-dependent in a universal way.  

Although the relation between the conformal frames has been understood in general, it is, perhaps, worth giving it a closer look in the concrete theory of fundamental interactions we are working with. This is the main subject of the present paper, in which we consider the Standard Model minimally coupled to $f (R)$ theory of gravity.  In addition, working in the Einstein frame, we are going to estimate potential observable effects of $f (R)$ gravity in a late-time universe in which the oscillating scalaron plays the role of dark matter~\cite{Cembranos:2008gj, Shtanov:2021uif}.

The Lagrangian of the Standard Model explicitly contains only one dimensionful parameter (in the units $\hbar = c = 1$), the vacuum expectation value $\eta_0$ of the Higgs field. However, according to our current understanding, this Lagrangian is to be regarded as an asymptotic local limit of the renormalisation-group flow. The physics that this theory describes at finite energy scales is characterised also by physical constants arising from the phenomenon known as dimensional transmutation. Most important of these is the parameter $\Lambda_\text{QCD}$, which enters the law of renormalisation of the coupling constant of strong interactions. Simplifying things by disregarding other gauge groups, we may take that $\Lambda_\text{QCD}$ and $\eta_0$ determine the masses of all observable particles: hadrons, leptons, and gauge~bosons. 

We will see that the scalaron in the Einstein frame couples to the Higgs field, in particular, through the scalaron-dependence of the new vacuum expectation value $\widetilde \eta (\phi)$. In this sense, the Einstein frame becomes the frame with transformed, or running mass units (in the terminology of~\cite{Dicke:1961gz, Faraoni:2006fx}) for all bare masses of elementary particles, which arise due to the Higgs mechanism. As regards the masses of bound states such as hadrons, they depend in a non-trivial way also on the implicit dimensionful quantum parameters such as $\Lambda_\text{QCD}$.  This opens up two possibilities of interpreting the theory in the Einstein frame: as a theory with running implicit parameters (which then becomes the frame with running mass units), equivalent to the theory in the original Jordan frame, or as a theory with fixed implicit parameters.  Although the two interpretations, in general, differ in their observable predictions, they become equivalent in situations with a completely stabilised or weakly excited scalaron.  The last situation arises in a late-time universe in which the excited scalaron plays the role of dark matter~\cite{Cembranos:2008gj, Shtanov:2021uif}, and we will show that the difference between these interpretations is practically negligible in this case. 

\section{Gravitational Constant in the Einstein Frame}

In this section, we review the well-known transition from the Jordan frame to the Einstein frame in the gravity sector. Our attention will be focused on the origin of the gravitational constant (Planck mass) in this frame.

Adopting the metric signature $(-, +, +, +)$, we write the gravitational action in the~form:
\begin{equation} \label{Sg}
S_g = \int d^4 x \sqrt{-g}\, f (R) \, .
\end{equation}

Note that the function $f (R)$ of the scalar curvature has mass dimension four in the unit system $ \hbar = c = 1$.  In general relativity, we have:
\begin{equation}\label{GR}
f_\text{GR} (R) = \frac{M_\text{P}^2}{3} \left( R - 2 \Lambda \right) \, , 
\end{equation}
where $M_\text{P} = \sqrt{3/16 \pi G} \approx 3 \times 10^{18}~\text{GeV}$ is a conveniently normalised Planck mass (the reason for our choice of this normalisation will become clear below). It is customary to explicitly introduce the factor ${M_\text{P}^2}/{3}$ in $f (R)$.  This can be done without loss of generality, but we will not do this here for better clarity.  The point is that any other constant of the same dimension can be factored out here, while our aim is to trace the origin of the physical Planck mass in the Einstein frame.

Proceeding to the Einstein frame, as a first step, one writes action~\eqref{Sg} in the form
\begin{equation}\label{Sg1}
S_g = \int d^4 x \sqrt{-g}\, \bigl[ \Omega R - h (\Omega) \bigr] \, ,
\end{equation}
where $\Omega$ is a new field with mass dimension two, and $h (\Omega)$ is the Legendre transform of $f (R)$.  It is defined by the following equations:
\begin{align} \label{leg1}
f' (R) &= \Omega \quad \Rightarrow \quad R = R (\Omega) \, , \\
h (\Omega) &= \bigl[ \Omega R - f (R) \bigr]_{R = R (\Omega)} \, . \label{leg2}
\end{align}

The inverse transform allows one to calculate $f (R)$ given $h (\Omega)$; it is obtained by variation of~\eqref{Sg1} with respect to $\Omega$:
\begin{align} \label{ileg1}
h' (\Omega) &= R \quad \Rightarrow \quad \Omega = \Omega (R) \, , \\
f (R) &= \bigl[ \Omega R - h (\Omega) \bigr]_{\Omega = \Omega (R)} \, . \label{ileg2}
\end{align}

These transformations may involve subtleties as to which solution is to be chosen in~\eqref{leg1} and~\eqref{ileg1}. Solutions of these equations are unique for convex functions, e.g., if $f'' (R) > 0$ everywhere in the domain of validity. In what follows, we will assume $\Omega$ to be positive; therefore, in view of~\eqref{leg1}, we also require\endnote{This requirement ensures that the effective gravitational coupling is positive in the Jordan frame, see~\cite{Sotiriou:2008rp, DeFelice:2010aj, Nojiri:2017ncd} for reviews.} $f' (R) > 0$.  

Our next step is to transform action~\eqref{Sg1} so that its  linear term %Please check that the intended meaning is retained. 
in $R$ takes the Einstein form. For this purpose, we perform a conformal transformation of the metric:
\begin{equation} \label{conform}
g_{\mu \nu} = \frac{M^2}{3 \Omega} \widetilde g_{\mu\nu} \, .
\end{equation}

Here, we explicitly introduced an arbitrary mass parameter $M$, compensating for the dimension of $\Omega$ so that the conformal transformation parameter is dimensionless (leaving the dimension of the metric intact).  With this transformation, we have:
\begin{equation}
\sqrt{- g}\, \Omega R = \frac{M^2}{3} \sqrt{- g}\, \frac{3 \Omega }{M^2}  R = \frac{M^2}{3} \sqrt{- \widetilde g}\, \left[ \widetilde R - \frac32 \left( \widetilde \nabla \ln \Omega \right)^2 + 3\, \widetilde \Box \ln \Omega \right] \, ,
\end{equation}
in which all objects related to the new metric $\widetilde g_{\mu\nu}$ are denoted by tildes. The last term is the total derivative and can be dropped. The transformed action~\eqref{Sg1} then becomes:
\begin{equation}\label{Sg2}
S_g = \int d^4 x \sqrt{- \widetilde g}\, \left[ \frac{M^2}{3} \widetilde R - \frac{M^2}{2} \left( \widetilde \nabla \ln \Omega \right)^2 - W (\Omega) \right] \, ,
\end{equation}
where
\begin{equation}\label{W}
W (\Omega) = \frac{M^4}{9} \frac{h (\Omega)}{\Omega^2} \, .
\end{equation}

We have obtained an Einstein theory of gravity with a minimally coupled scalar field and with a Planck mass $M$.  The theory is stable only if the potential $W (\Omega)$ has a minimum at $\Omega = \Omega_0$. In view of system~\eqref{leg1}--\eqref{ileg2}, this condition is equivalent to the existence of $R_0$, such that:
\begin{equation}\label{stab}
R_0 f' (R_0) = 2 f (R_0) \, , \qquad \frac{1}{f'' (R_0)} - \frac{R_0^2}{2 f (R_0)} > 0 \, .
\end{equation}

Then, $\Omega_0 = f' (R_0)$.  From these relations, it is clear that the values of $\Omega_0$ and $R_0$ are independent of $M$ introduced in~\eqref{conform}.  This is also evident from the fact that the parameter $M$ enters only as an overall scaling in potential~\eqref{W}. 

We then introduce a scalar field (scalaron) $\phi$ with a canonical kinetic term and with a minimum of potential at $\phi = 0$ by setting:\endnote{If $W (\Omega)$ does not have a minimum, we still can write~\eqref{phi} by choosing $\Omega_0$ arbitrarily.  In this case, however, the theory will not have a stable point.}
\begin{equation} \label{phi}
\Omega = \Omega (\phi) = \Omega_0 e^{\phi / M} \, .
\end{equation} 

Action~\eqref{Sg2}, eventually, becomes:
\begin{equation}\label{Sg3}
S_g =  \int d^4 x \sqrt{- \widetilde g}\, \left[ \frac{M^2}{3} \widetilde R - \frac12 \left( \widetilde \nabla \phi \right)^2 - V (\phi) \right] \, ,
\end{equation}
where the scalaron potential $V (\phi)$ is calculated by using~\eqref{leg2},~\eqref{W} and~\eqref{phi}:
\begin{equation}\label{V}
V (\phi) \equiv W \left( \Omega (\phi ) \right) = \frac{M^4}{9} \left[ \frac{R}{\Omega} - \frac{f (R)}{\Omega^2} \right]_{\substack{R = R (\Omega) \\ \Omega = \Omega (\phi)}} \, .
\end{equation}

We note that the mass parameter $M$ has appeared as the Planck mass in action~\eqref{Sg3}, and as an overall scaling in potential~\eqref{V}.\endnote{The appearance of a simple expression $\phi / M$ (without numerical factor)  in the exponent~\eqref{phi} and in all subsequent exponents of this type is the reason why we have chosen the particular normalisation constant in~\eqref{conform} and in~\eqref{GR}.} However, this parameter was introduced in~\eqref{conform} quite arbitrarily.  This might appear paradoxical, as if $f (R)$ gravity does not predict a specific value for the gravitational constant in the Einstein frame. This paradox is resolved by examining the matter part of the action, which  is done in the following section.

\section{Coupling of Gravity to Matter}
\label{sec:matter}

As regards the matter action in the Jordan frame, we take it to be that of the Standard Model minimally coupled to gravity. Proceeding to the Einstein frame via~\eqref{conform} affects this action as well. Note, however, that most of the Standard Model action is classically conformally invariant (with proper conformal transformation of the matter fields), and, therefore, will retain its original form after transformation~\eqref{conform}. The only part that breaks conformal invariance is the Higgs sector, with the action:
\begin{equation}\label{Sh}
S_\text{H} = - \int d^4 x \sqrt{-g} \left[ g^{\mu\nu} \left( D_\mu \Phi \right)^\dagger D_\nu \Phi + \lambda \left( \Phi^\dagger \Phi - \frac{\eta_0^2}{2} \right)^2 \right] \, .
\end{equation}

Here, $D_\mu$ is the gauge covariant derivative involving the SU(2) and U(1) electroweak gauge fields and acting on the Higgs doublet $\Phi$, and
$\eta_0$ is the symmetry-breaking parameter in the Jordan frame. After the conformal transformation~\eqref{conform},~\eqref{phi} is accompanied by the transformation of the Higgs scalar field:
\begin{equation}\label{Htrans}
\Phi = \frac{\sqrt{3 \Omega}}{M}\, \widetilde \Phi \, ,
\end{equation}
and this action becomes:\endnote{A similar result of the conformal transformation of fields was under consideration, e.g., in~\cite{Rudenok:2014daa, Burrage:2018dvt, Copeland:2021qby}. In passing, we note that conformal transformation, in our theory, where the Higgs scalar field is minimally coupled to gravity, looks much simpler compared to the case of its non-minimal coupling, as is the case, e.g., in the model of Higgs inflation~\cite{Bezrukov:2007ep}.}
%\end{paracol}
%\nointerlineskip
\begin{align}\label{Shn}
S_\text{H} = - \int d^4 x \sqrt{- \widetilde g}\, \widetilde g^{\mu\nu} \left[ \left( D_\mu \widetilde \Phi \right)^\dagger D_\nu \widetilde \Phi + \frac{1}{2 M} \widetilde \nabla_\mu  \left( \widetilde \Phi^\dagger \widetilde \Phi \right) \widetilde \nabla_\nu \phi + \frac{1}{4 M^2} \widetilde \Phi^\dagger \widetilde \Phi\, \widetilde \nabla_\mu \phi \widetilde \nabla_\nu \phi \right] \nonumber \\ - \lambda \int d^4 x \sqrt{- \widetilde g}  \left( \widetilde \Phi^\dagger \widetilde \Phi - \beta e^{- \phi / M} \frac{\eta_0^2}{2} \right)^2 \, , 
\end{align}
%\begin{paracol}{2}
%\linenumbers
%\switchcolumn
where
\begin{equation}\label{beta}
\beta = \frac{M^2}{3 \Omega_0} > 0
\end{equation}
is a dimensionless constant. We observe the appearance of non-renormalisable interactions of the scalaron $\phi$ with the Higgs field in~\eqref{Shn}, which, however, are all suppressed by inverse powers of the Planck mass $M$. 

From~\eqref{Shn}, one observes that the Higgs vacuum expectation value in the Einstein frame~is: 
\begin{equation} \label{vscale}
\widetilde \eta (\phi)  = \sqrt{\beta}\, e^{- \phi / 2M} \eta_0 \, .
\end{equation}

It is this parameter that will determine the bare masses of all fermions and gauge bosons in the model, which are all proportional to it. Now, in the scalaron vacuum $\phi = 0$, the ratio of $\widetilde \eta_0 = \widetilde \eta (0) = \sqrt{\beta} \eta_0$ to the Planck mass $M$ in~\eqref{Sg3} is:
\begin{equation}
\frac{\widetilde \eta_0}{M} = \frac{\eta_0}{\sqrt{3\, \Omega_0}} \, .
\end{equation}

This ratio is independent of the chosen scale $M$ in~\eqref{conform}, and is uniquely determined by the original Jordan-frame actions~\eqref{Sg} and~\eqref{Sh}. Only this ratio makes sense and is physically measurable.  This explains the freedom of choosing $M$ arbitrarily in~\eqref{conform}. Assigning the observed value $\widetilde \eta_0 \approx 246~\text{GeV}$, we should then equate $M$ to the Planck mass \linebreak $M_\text{P} = \sqrt{3/16 \pi G} \approx 3 \times 10^{18}~\text{GeV}$. 

The same reasoning applies to the issue of cosmological constant in the Einstein frame.  We observe that its purely gravitational contribution $\widetilde \Lambda$ in this frame is given by the minimum of the scalaron potential~\eqref{V}: 
\begin{equation}
\widetilde \Lambda = \frac{M^2}{24} \frac{R_0^2}{f (R_0)} \, ,
\end{equation}
where we have used equation~\eqref{stab}. We see that the observed ratio $\widetilde \Lambda / M^2$ is also independent of the chosen scale $M$ in~\eqref{conform}.  The mass hierarchy problem of modern cosmology can be expressed as:
\begin{equation}
\frac{\widetilde \Lambda}{M^2} \ll \frac{\widetilde \eta_0^2}{M^2} \ll 1 \quad \Rightarrow \quad \frac{R_0^2}{f (R_0)} \ll \frac{R_0 \eta_0^2}{f (R_0)} \ll 1 \, ,
\end{equation}
where, we remember, $R_0$ is a solution of~\eqref{stab}. The last set of inequalities are written in terms of the action in the original Jordan frame.

This analysis would be the whole story for a world described by classical fields. However, the fields in the Standard Model Lagrangian are quantum, and their quantum dynamics are non-trivial.

The Lagrangian of the Standard Model in the Jordan frame contains only one explicit dimensionful parameter, the vacuum expectation value $\eta_0$ of the Higgs field. However, according to modern understanding, this Lagrangian is to be regarded as the relevant part of the low-energy (or large-scale) action of some renormalisation-group flow (see, e.g.,~\cite{Costello}). The physics that this theory describes at finite energy scales is also characterised by implicit dimensionful parameters arising in what is known as dimensional transmutation. Such is the QCD parameter $\Lambda_\text{QCD}$ that enters the law of renormalisation-group flow of the coupling constant of strong interactions and determines the masses of hadrons (see, e.g.,~\cite{Chivukula:2004hw, Deur:2014qfa}). Simplifying the situation by disregarding other gauge interactions, we may take that two dimensionful parameters, $\Lambda_\text{QCD}$ and $\eta_0$, control the masses of all particles and bound states, including hadrons, in the Jordan frame.  With this simplification, we will have, for the \emph{i}th particle mass: 
\begin{equation} \label{mi}
m_i = \Lambda_\text{QCD}\, f_i \left( \frac{\eta_0}{\Lambda_\text{QCD}} \right) \, ,
\end{equation} 
where $f_i (x)$ are some dimensionless functions.

This consideration opens up two possibilities of interpreting the theory in the Einstein~frame.

\subsection{Einstein Frame with Running Implicit Parameters}
\label{sec:running}

In the matter Lagrangian density, we can proceed to the Einstein frame by the conformal transformation~\eqref{conform} of the metric, scaling the Higgs field and the fermionic fields $\psi$~accordingly: 
\begin{equation} \label{Fconform}
{\cal L} = {\cal L} \left( \beta e^{-\phi/M} \widetilde g_{\mu\nu}, \, \beta^{-1/2} e^{\phi / 2 M} \widetilde \Phi, \, \beta^{-3/4} e^{3 \phi / 4 M} \widetilde  \psi \right) \equiv \widetilde {\cal L} \left( \widetilde g_{\mu\nu}, \, \widetilde \Phi , \, \widetilde \psi , \, \phi \right) \, .
\end{equation}

Here, by virtue of an almost perfect conformal invariance of the action, the last expression differs from the Lagrangian density ${\cal L} \left( \widetilde g_{\mu\nu}, \, \widetilde \Phi , \, \widetilde \psi \right)$ only in the Higgs part~\eqref{Shn}.  However, the scalaron-dependent scaling of quantum fields, together with~\eqref{conform}, lead us to a quantum theory with accordingly scaled, implicit quantum dimensionful parameters. In particular, the QCD parameter in the Einstein frame is locally scaled as:
\begin{equation} \label{Lscale}
\widetilde \Lambda_\text{QCD} (\phi) = \frac{M}{\sqrt{3 \Omega}} \Lambda_\text{QCD} = \sqrt{\beta}\, e^{-\phi / 2 M}  \Lambda_\text{QCD} \, ,
\end{equation}
similarly to the scaling~\eqref{vscale} of the parameter $\eta_0$ in~\eqref{Shn}.  Replacing $\Lambda_\text{QCD}$ and $\eta_0$ in~\eqref{mi} with their scaled values $\widetilde \Lambda_\text{QCD} (\phi)$ and $\widetilde \eta (\phi)$ in the Einstein frame, we observe that all masses are scaled in the same way:
 \begin{equation} \label{mscale}
\widetilde m_i = \sqrt{\beta}\, e^{-\phi / 2 M}  m_i \, .
\end{equation}

Their ratios to the Planck mass $M$ at the scalaron vacuum $\phi = 0$ are, again, independent of our choice of $M$ in~\eqref{conform}.  

Scaling~\eqref{mscale} describes the situation of an {\em Einstein frame with running units\/}, an option first discussed in~\cite{Dicke:1961gz} and further elucidated in~\cite{Faraoni:2006fx}.  Here, the scalaron, in addition to the interaction with the Higgs field through the explicit mass parameter $\widetilde \eta (\phi)$ in~\eqref{Shn}, also interacts with matter implicitly via~\eqref{Lscale} and, therefore,~\eqref{mscale}. 
%(For $|\phi| / M \ll 1$, however, the scalaron coupling to matter is very weak.)  
One can arrive at the same picture by considering the matter Lagrangian density in the Einstein frame with transformed metric field only, i.e.,
\begin{equation}\label{running}
{\cal L} = {\cal L} \left( \beta e^{-\phi/M} \widetilde g_{\mu\nu}, \, \Phi, \, \psi \right) \, ,
\end{equation}
and treating $\Lambda_\text{QCD}$ and related quantum condensates of the $\psi$ fields as quantities coinciding with their values in the Jordan frame. In this case, one obtains the usual relation between the stress--energy tensors in two frames:
\begin{equation}
\widetilde T_{\mu\nu} = \beta e^{- \phi / M} T_{\mu\nu} \, .
\end{equation}

An expression for the mass of a static localised object is:
\begin{equation}
m_i = \int_\Sigma T_{\mu \nu} n^\mu \xi^\nu d \mu_\Sigma \, ,
\end{equation}
where the integral is taken over a hypersurface $\Sigma$, with $n^\mu$ being the vector field of unit normal to this hypersurface; $\xi^\mu$ is the timelike killing vector field such that $\xi^\mu \xi_\mu = -1$ at spatial infinity, and $d \mu_\Sigma$ is the volume measure determined by the induced metric on $\Sigma$. Using the conformal transformation laws of these quantities, we again arrive at the transformation law~\eqref{mscale} for the mass. 

The two (Jordan and Einstein) frames in this interpretation are equivalent; it is a matter of convenience in regards to which frame one chooses to work with, as long as one keeps track of scaling~\eqref{Lscale} and~\eqref{mscale}. In particular, all objects (massive as well as massless) move along the geodesics of the Jordan-frame metric $g_{\mu\nu} = \beta e^{-\phi/M} \widetilde g_{\mu\nu}$, which is the ``observable'' metric in all respects~\cite{Dicke:1961gz, Faraoni:2006fx}.\endnote{In the context of scalar-tensor theories, such an equivalence between frames on the tree level was recently demonstrated in~\cite{Copeland:2021qby}. On the one-loop level, on-shell equivalence between conformal frames was demonstrated previously in~\cite{Kamenshchik:2014waa, Ruf:2017xon}.}

\subsection{Einstein Frame with Fixed Implicit Parameters}

Treating the fields in the Lagrangian $\widetilde {\cal L} \left( \widetilde g_{\mu\nu}, \, \widetilde \Phi , \, \widetilde \psi , \, \phi \right)$ of~\eqref{Fconform} as given, and ``forgetting'' about their Jordan-frame origin, one can specify their quantum theory by a $\phi$-independent implicit quantum parameter $\Lambda_\text{QCD}$ in the Einstein frame. This will create a situation quite different from that of Section~\ref{sec:running}, since now, the bare masses of quarks, leptons, and gauge fields will depend on the scalaron, as before, through the Higgs expectation-value parameter~\eqref{vscale}, while the hadron masses will be given by:
\begin{equation}\label{mrescale}
\widetilde m_i = \Lambda_\text{QCD}\, f_i \left( \frac{\widetilde \eta (\phi) }{\Lambda_\text{QCD}} \right) \, ,
\end{equation}
depending on the scalaron in a way that is more complicated than~\eqref{mscale}.   We see that, in our framework and strictly speaking, there is no conformal frame with completely fixed units. Particle masses in this frame depend on the scalaron field in a different manner.

For the scalaron in the vacuum, this difference between frames will not be revealed, but it will exist in situations where the scalaron is dynamically excited.  One such situation is considered in Section~\ref{sec:DM}.  

\subsection{Conformal Anomaly}

The quantum loop corrections to the classical action lead to the effect that the extra degree of freedom present in $f (R)$ gravity couples to matter also due to the conformal anomaly.  This is most easily seen in the Einstein frame, in which couplings between the scalaron and gauge fields with strength tensor $F_{\mu\nu}$, for small values of $|\phi|/M$, are proportional to ~\cite{Cembranos:2008gj, Katsuragawa:2016yir}:
\begin{equation} \label{anom}
\alpha_\text{g}  \frac{\phi}{ M} \, \text{tr}\, F_{\mu\nu} F^{\mu\nu} \, , 
\end{equation}
where $\alpha_\text{g} = e_\text{g}^2/4 \pi$, and $e_\text{g}$ is the relevant gauge-coupling parameter. 

\section{Light Scalaron as Dark Matter}
\label{sec:DM}

Since the original value of $M$ can be fixed arbitrarily (as was shown Section~\ref{sec:matter}), we fix it in what follows so that $\beta = 1$ (this constant is defined in~\eqref{beta}).

The $\phi$-dependence of the particle masses, such as~\eqref{mscale} or~\eqref{mrescale}, might be potentially interesting in the case of a classically evolving scalaron. This may lead to important phenomena in the early universe, where the scalaron can be highly excited.  For example, if the scalaron plays the role of an inflaton, when proceeding to the Einstein frame, it may be necessary to take into account the dependence~\eqref{mscale} of the masses of fundamental particles on the scalaron field during inflation.  The same dependence is responsible for particle creation during preheating in such an inflationary theory~\cite{Rudenok:2014daa, Gorbunov:2010bn}.

In this paper, however, we will focus on the situation that arises in a late-time universe in which the oscillating scalaron plays the role of dark matter~\cite{Cembranos:2008gj, Shtanov:2021uif}. In this case, the scalaron oscillations might lead to potentially observable effects. Such effects are determined by the small ratio:\endnote{For the solar neighbourhood, the dark-matter density is $\rho_\text{dm} \simeq 10^{-2} M_\odot / \text{pc}^3$~\cite{Read:2014qva}, which gives $\left( {\rho^{}_\text{s}}/{\overline \rho_\text{s}} \right)^{1/2} \simeq 500$.}
\begin{equation}\label{phitoM}
\frac{|\phi|}{M} \simeq \frac{\sqrt{2 \rho_\text{s}}}{M^2} \frac{M}{m} \simeq 10^{- 33} \left( \frac{\rho_\text{s}}{\overline \rho_\text{s}} \right)^{1/2} \left( 1 + z \right)^{3/2}~\frac{\text{eV}}{m} \, ,
\end{equation}
where $z$ is the cosmological redshift, $\rho_\text{s}$ is the local scalaron energy density, $\overline \rho_\text{s}$ is its spatial average in the universe, and $m$ is the scalaron mass. The effects are very small because the scalaron mass in the interpretation of the Einstein frame with running units is bounded from below by non-observation~\cite{Kapner:2006si, Adelberger:2006dh} of the additional Yukawa forces~\cite{Stelle:1977ry} between non-relativistic masses (see also~\cite{Cembranos:2008gj, Perivolaropoulos:2019vkb}):
\begin{equation}
m \geq 2.7 \times 10^{-3}\, \text{eV} \quad \text{at 95\% C.L.}
\end{equation}

As regards the Einstein frame with constant implicit parameter $\Lambda_\text{QCD}$, the $\phi$-dependence in~\eqref{mrescale} will have additional smallness because the contribution of the bare quark masses to the masses of hadrons constitute only a tiny fraction of the total mass~\cite{Chivukula:2004hw}.

Let us examine this in more detail for the late-time universe. The effect is most prominent in the interpretation of the Einstein frame with running units, where the observable metric will be that of the Jordan frame. Then, in the Einstein frame, all masses scale with the scalaron as~\eqref{mscale}. This produces second-order effects in the small gravitational potential because of the smallness of the ratio~\eqref{phitoM}. However, in the observable metric, this gives an effect of the first order because of the scaling $g_{\mu\nu} = e^{-\phi/M} \widetilde g_{\mu\nu}$.  This will not affect null geodesics apart from additional redshift, but will produce an effective Newtonian potential,
\begin{equation}\label{Newpot}
\varphi_\text{eff} = \varphi - \frac{\phi}{2 M} ,
\end{equation}
for non-relativistic matter. Here, $\varphi$ is the Newtonian potential in the Einstein-frame metric $\widetilde g_{\mu\nu}$, which is determined by the distribution of the usual matter and dark matter in the form of scalaron. In the Newtonian approximation to gravity, which is of relevance here, it is {described} by the Poisson equation ${\boldsymbol \nabla}^2 \varphi = 3 \rho / 4 M^2 $, where $\rho$ is the total energy density of matter, including the scalaron field.

The scalaron field $\phi$ of mass $m$ is oscillating in time with the following period:
\begin{equation}\label{period}
t = \frac{2 \pi}{m} \approx 4.1 \times 10^{-15} \left( \frac{\text{eV}}{m} \right)~\text{s} \, . 
\end{equation}

Therefore, {the last term in the effective Newtonian potential~\eqref{Newpot} is rapidly oscillating in time with space-dependent amplitude. Since one is interested in the motion of astrophysical bodies on time scales much longer than~\eqref{period}, one should perform time averaging of their dynamics in such a rapidly oscillating potential.} The solution to this {classical} problem is well known (see \S30 in~\cite{LL}). The average {long-term} effective potential acting on test bodies is given by:
\begin{equation}
\overline \varphi_\text{eff} = \varphi + \frac{\overline{ \left( {\boldsymbol \nabla} \phi \right)^2 }}{8 m^2 M^2} = \varphi + \frac{\left( {\boldsymbol \nabla} \phi_0 \right)^2}{16 m^2 M^2} \, ,
\end{equation}
where overline denotes averaging over a period of oscillations of $\phi$, and $\phi_0$ is the space-dependent amplitude of its oscillations.  Since the scalaron energy density $\rho_\text{s} = \frac12 m^2 \phi_0^2$, we~have:
\begin{equation} \label{phieff}
\overline \varphi_\text{eff} = \varphi + \frac{\left( {\boldsymbol \nabla} \rho_\text{s} \right)^2}{32 m^4 M^2 \rho_\text{s}} = \varphi + \frac{\left( {\boldsymbol \nabla} {\boldsymbol \nabla}^2 \varphi_\text{s} \right)^2}{24 m^4 {\boldsymbol \nabla}^2 \varphi_\text{s}} \, ,
\end{equation}
where we have used the Poisson equation $\rho_\text{s} = \frac43 M^2 {\boldsymbol \nabla}^2 \varphi_\text{s}$ for the scalaron contribution $\varphi_\text{s}$  to the total gravitational potential $\varphi$. 

The scalaron energy density, and the gravitational potential $\varphi_\text{s}$, hence, varies on the spatial scale of the de~Broglie wavelength  (see~\cite{Hui:2021tkt} for a review on such wave dark matter):
\begin{equation} \label{dB}
\lambda_\text{dB} \simeq  \frac{2 \pi}{m v} = 124\, \left( \frac{10^{-3}~\text{eV}}{m} \right)~\left( \frac{10^{-3} }{v} \right)~\text{cm} \, ,
\end{equation}
where $v$ is the velocity dispersion in a virialised dark-matter halo (in units of the speed of light).  For the last term in~\eqref{phieff}, this gives an estimate:
\begin{equation} 
\frac{\left( {\boldsymbol \nabla} {\boldsymbol \nabla}^2 \varphi_\text{s} \right)^2}{24 m^4 {\boldsymbol \nabla}^2 \varphi_\text{s}} \sim \frac{ | \varphi_\text{s} |}{24 m^4 \lambda_\text{dB}^4} \sim 10^{-4} v^4 | \varphi_\text{s} | \, .
\end{equation}

Here, we have replaced all spatial gradients by the characteristic inverse length $\lambda_\text{dB}^{-1}$.  For typical velocity dispersions $v\sim10^{-2}$$-$$10^{-3}$, this is many orders of magnitude smaller than $| \varphi_\text{s} | $. The de~Broglie wavelength scale~\eqref{dB} and time scale $t_\text{dB} = \lambda_\text{dB} / v$ themselves are rather small for masses $m \gtrsim 10^{-3}$\,eV allowable in this theory.  The direct effects of the scalaron oscillations are, thus, quite negligible, and the Jordan and Einstein conformal frames are practically indistinguishable in this case.

Couplings~\eqref{anom} due to the conformal anomaly lead to the scalaron decay into photons, with lifetime $\tau \sim M^2 / \alpha^2 m^3 \sim 10^{36}\, (\text{eV}/m)^3\, \text{yr}$, exceeding the age of the universe\linebreak  ($1.4 \times 10^{10}\, \text{yr}$) for $m \ll 10^8$\,eV.  (Here, $\alpha$ is the fine-structure constant.) For the scalaron masses of order $m\sim10^{-3}$\,eV, as in the scenario of~\cite{Shtanov:2021uif}, such a light scalaron dark matter appears to be quite ``sterile'' and hard to detect by means other than gravitationally. The smallness of the specific gravitational manifestations in the scenario under consideration would make it very difficult to establish that we are dealing with $f (R)$ gravity. Perhaps, this could be done only by detecting a specific Yukawa contribution to gravitational forces at submillimetre spatial scales~\cite{Kapner:2006si, Adelberger:2006dh, Stelle:1977ry, Perivolaropoulos:2019vkb}.

%%%%%%%%%%%%%%%%%%%%%%%%%%%%%%%%%%%%%%%%%%
\section{Discussion}

A generic gravity theory with action~\eqref{Sg} has a stable point if there exists a solution to~\eqref{stab}.  In this case, it is conformally transformed by~\eqref{conform} to a general-relativistic theory~\eqref{Sg3} with a minimally coupled scalar field with potential~\eqref{V} that has minimum at $\phi = 0$. The gravitational coupling at this stage is introduced arbitrarily in~\eqref{conform}.  However, its physical value in the Einstein frame is uniquely determined by the function $f (R)$, together with the matter action in the Jordan frame. This was demonstrated in Section~\ref{sec:matter}, where we considered the action of the Standard Model minimally coupled to gravity in the Jordan frame.

We reviewed and further elucidated two possible interpretations of this model in the Einstein frame, namely, as the frame with either running or constant implicit quantum parameters. The Standard Model contains at least one such implicit parameter $\Lambda_\text{QCD}$, which transforms under conformal transformation of the metric. The difference between these two interpretations, however, is practically insignificant in situations where the scalaron field is weakly excited.  We verified this for the case of cosmology in which a light scalaron plays the role of dark matter.  In such a scenario, the fact that we are dealing with $f (R)$ gravity could be observationally verified, perhaps, only by detecting a specific Yukawa contribution to gravitational forces at submillimetre spatial scales~\cite{Kapner:2006si, Adelberger:2006dh, Stelle:1977ry, Perivolaropoulos:2019vkb}.

%%%%%%%%%%%%%%%%%%%%%%%%%%%%%%%%%%%%%%%%%%
%\section{Conclusions}
%
%This section is not mandatory, but can be added to the manuscript if the discussion is unusually long or complex.

%%%%%%%%%%%%%%%%%%%%%%%%%%%%%%%%%%%%%%%%%%
\vspace{6pt} 

%%%%%%%%%%%%%%%%%%%%%%%%%%%%%%%%%%%%%%%%%%
%% optional
%\supplementary{The following are available online at \linksupplementary{s1}, Figure S1: title, Table S1: title, Video S1: title.}

% Only for the journal Methods and Protocols:
% If you wish to submit a video article, please do so with any other supplementary material.
% \supplementary{The following are available at \linksupplementary{s1}, Figure S1: title, Table S1: title, Video S1: title. A supporting video article is available at doi: link.} 

%%%%%%%%%%%%%%%%%%%%%%%%%%%%%%%%%%%%%%%%%%
%\authorcontributions{For research articles with several authors, a short paragraph specifying their individual contributions must be provided. The following statements should be used ``Conceptualization, X.X. and Y.Y.; methodology, X.X.; software, X.X.; validation, X.X., Y.Y. and Z.Z.; formal analysis, X.X.; investigation, X.X.; resources, X.X.; data curation, X.X.; writing---original draft preparation, X.X.; writing---review and editing, X.X.; visualization, X.X.; supervision, X.X.; project administration, X.X.; funding acquisition, Y.Y. All authors have read and agreed to the published version of the manuscript.'', please turn to the  \href{http://img.mdpi.org/data/contributor-role-instruction.pdf}{CRediT taxonomy} for the term explanation. Authorship must be limited to those who have contributed substantially to the work~reported.}

\funding{This research was funded by the National Academy of Sciences of Ukraine project number {0121U109612.}}

\institutionalreview{Not applicable. %In this section, please add the Institutional Review Board Statement and approval number for studies involving humans or animals. Please note that the Editorial Office might ask you for further information. Please add ``The study was conducted according to the guidelines of the Declaration of Helsinki, and approved by the Institutional Review Board (or Ethics Committee) of NAME OF INSTITUTE (protocol code XXX and date of approval).'' OR ``Ethical review and approval were waived for this study, due to REASON (please provide a detailed justification).'' OR ``Not applicable'' for studies not involving humans or animals. You might also choose to exclude this statement if the study did not involve humans or animals.
}

\informedconsent{Not applicable. %Any research article describing a study involving humans should contain this statement. Please add ``Informed consent was obtained from all subjects involved in the study.'' OR ``Patient consent was waived due to REASON (please provide a detailed justification).'' OR ``Not applicable'' for studies not involving humans. You might also choose to exclude this statement if the study did not involve humans.

%Written informed consent for publication must be obtained from participating patients who can be identified (including by the patients themselves). Please state ``Written informed consent has been obtained from the patient(s) to publish this paper'' if applicable.
}

\dataavailability{No new data were created or analysed in this study. Data sharing is not applicable to this article. %In this section, please provide details regarding where data supporting reported results can be found, including links to publicly archived datasets analyzed or generated during the study. Please refer to suggested Data Availability Statements in section ``MDPI Research Data Policies'' at \url{https://www.mdpi.com/ethics}. You might choose to exclude this statement if the study did not report any data.
} 

\acknowledgments{\textls[-15]{The author is grateful to Maria Khelashvili and Anton Rudakovskyi for discussion.}}

\conflictsofinterest{%Declare conflicts of interest or state ``
The author declares no conflict of interest.
%'' Authors must identify and declare any personal circumstances or interest that may be perceived as inappropriately influencing the representation or interpretation of reported research results. Any role of the funders in the design of the study; in the collection, analyses or interpretation of data; in the writing of the manuscript, or in the decision to publish the results must be declared in this section. If there is no role, please state ``
The funders had no role in the design of the study; in the collection, analyses, or interpretation of data; in the writing of the manuscript, or in the decision to publish the results. %''.
} 

%% Optional
%\sampleavailability{Samples of the compounds ... are available from the authors.}

%%%%%%%%%%%%%%%%%%%%%%%%%%%%%%%%%%%%%%%%%%
%% Only for journal Encyclopedia
%\entrylink{The Link to this entry published on the encyclopedia platform.}

%%%%%%%%%%%%%%%%%%%%%%%%%%%%%%%%%%%%%%%%%%
%% Optional
\abbreviations{Abbreviations}{

The following abbreviations are used in this manuscript:\\

\noindent 
\begin{tabular}{@{}ll}
C.L. & Confidence Level \\
GR & General Relativity \\
QCD & Quantum Chromodynamics
%DOAJ & Directory of open access journals\\
%TLA & Three letter acronym\\
%LD & linear dichroism
\end{tabular}}

\begin{adjustwidth}{-\extralength}{0cm}
%%%%%%%%%%%%%%%%%%%%%%%%%%%%%%%%%%%%%%%%%%
%\end{paracol}
%%%%%%%%%%%%%%%%%%%%%%%%%%%%%%%%%%%%%%%%%%
% To add notes in main text, please use \endnote{} and un-comment the codes below.
\begin{adjustwidth}{0cm}{0cm}
\printendnotes[custom]
\end{adjustwidth}

%\centering %% If there is a figure in wide page, please release command \centering

%%%%%%%%%%%%%%%%%%%%%%%%%%%%%%%%%%%%%%%%%%
\reftitle{References}

\end{adjustwidth}
% The following MDPI journals use author-date citation: Admsci,  Arts, Econometrics, Economies, Genealogy, Humanities, IJFS, Jintelligence, JRFM, Languages, Laws, Literature, Religions, Risks, Social Sciences. For those journals, please follow the formatting guidelines on http://www.mdpi.com/authors/references
% To cite two works by the same author: \citeauthor{ref-journal-1a} (\citeyear{ref-journal-1a}, \citeyear{ref-journal-1b}). This produces: Whittaker (1967, 1975)
% To cite two works by the same author with specific pages: \citeauthor{ref-journal-3a} (\citeyear{ref-journal-3a}, p. 328; \citeyear{ref-journal-3b}, p.475). This produces: Wong (1999, p. 328; 2000, p. 475)

%%%%%%%%%%%%%%%%%%%%%%%%%%%%%%%%%%%%%%%%%%
%% for journal Sci
%\reviewreports{\\
%Reviewer 1 comments and authors’ response\\
%Reviewer 2 comments and authors’ response\\
%Reviewer 3 comments and authors’ response
%}
%%%%%%%%%%%%%%%%%%%%%%%%%%%%%%%%%%%%%%%%%%
\end{document}